\documentclass[twocolumn]{aastex631}

\usepackage{CJK}
\usepackage{verbatim} 
\usepackage[T1]{fontenc}
\usepackage[utf8]{inputenc}
\usepackage{graphicx}
\usepackage{xcolor}
\usepackage{amsmath}
\usepackage{amssymb}
\usepackage{booktabs}
\usepackage{enumitem}
\usepackage{microtype}
\usepackage{tikz}

\begin{document}
\begin{CJK*}{UTF8}{gkai}
\title{ASKAP Discovery of a Giant Odd Radio Circle at $z=0.15$}

\author[0000-0002-2432-2587]{Renzhi Su (苏仁智)}
\email{rzsu.astro@gmail.com}
\affiliation{Shanghai Astronomical Observatory, Chinese Academy of Sciences, 80 Nandan Road, Shanghai 200030, China}

\author[0000-0002-4455-6946]{Minfeng Gu (顾敏峰)}
\email{gumf@shao.ac.cn}
\affiliation{Shanghai Astronomical Observatory, Chinese Academy of Sciences, 80 Nandan Road, Shanghai 200030, China}



\begin{abstract}
Odd radio circles (ORCs) are a recently recognized class of faint, roughly circular diffuse radio sources, typically a few hundred kiloparsecs across, whose origin is still debated. We report the serendipitous discovery of a giant ORC, ORC~J1928$-$5056, in 856~MHz continuum data from the ASKAP observations.  The emission forms a single, mildly elliptical structure $4^{\prime}$ across, corresponding to $0.63$~Mpc at the redshift $z=0.1507$ of the central galaxy LEDA~140850, the brightest cluster galaxy of Abell~3639, which makes it one of the largest single ORCs known. After subtracting the radio nucleus associated with the host galaxy, the ORC has a flux density of 73~mJy integrated within an area of $\sim 12$~arcmin$^2$, i.e.\ a mean surface brightness of only $1.7$~$\mu$Jy~arcsec$^{-2}$, and a rest-frame luminosity of $L_{1.4}=2\times10^{24}$~W~Hz$^{-1}$, comparable to those of the most luminous ORCs. Including its nucleus, the system reaches $L_{1.4}=6.6\times10^{24}$~W~Hz$^{-1}$, higher than that of any other ORC reported so far. The host is a massive ($M_{\ast}\sim10^{11}\,M_{\odot}$), passive elliptical galaxy with negligible star formation. It contains an AGN jet with a power-law spectrum of index $\alpha=-0.877\pm0.089$ between 856 and 1656~MHz. Finally, we discuss the implications for ORC formation models and the observations needed to establish the nature of this system.

\end{abstract}



\section{Introduction}
\label{sec:intro}

Odd radio circles (ORCs) are faint, roughly circular patches of diffuse radio emission discovered serendipitously in continuum images from the Australian Square Kilometre Array Pathfinder (ASKAP) Evolutionary Map of the Universe (EMU) pilot survey \citep{norris2021}; two further examples were identified, in archival Giant Metrewave Radio Telescope data \citep{norris2021} and in ASKAP imaging \citep{koribalski2021}. The sample has since grown thanks to deep MeerKAT observations \citep[e.g.][]{norris2025} and searches of the EMU data, both automated and visual \citep[e.g.][]{gupta2025,filipovic2026,su2026}. Related shell systems have been identified around interacting early-type galaxies and in merging groups \citep{bulbul2024,koribalski2024,koribalski2024b,koribalski2026}; similar rings have been reported at other radio frequencies \citep{kumari2024,hota2025}. Most confirmed ORCs are hosted by massive elliptical galaxies and span $\sim$100--500~kpc, with integrated flux densities of only a few mJy at GHz frequencies \citep{norris2021b,rupke2024}; their edges are usually brighter than their interiors.

The physical origin of ORCs remains unknown. Proposed mechanisms include spherical blast waves powered by cataclysmic events such as supernova or tidal disruption event in the host galaxy \citep{omar2022,omar2022b}, termination shocks of powerful starburst winds \citep{lochhaas2018,coil2024,fujita2024}, AGN jet-inflated bubbles viewed in projection \citep{lin2024}, faded radio-galaxy lobes re-energized by a passing shock (``phoenixes''; \citealp{shabala2024}), virial shocks around massive halos \citep{yamasaki2024}, and merger-driven shocks in galaxy groups and clusters \citep{dolag2023,ivleva2026,bulbul2024}. Distinguishing among these scenarios requires accurate sizes, luminosities, spectra, polarization properties, and host-galaxy properties.

In this paper, we report the discovery, in ASKAP FLASH continuum data, of ORC~J1928$-$5056, one of the largest ORCs known, centered near the luminous elliptical galaxy LEDA~140850, the brightest cluster galaxy (BCG) of Abell~3639. Section~\ref{sec:obs} describes the observations and data reduction, Section~\ref{sec:results} presents our results and discusses their implications, and Section~\ref{sec:summary} summarizes our conclusions. We adopt a flat $\Lambda$CDM cosmology with $\Omega_{\rm m}=0.3$ and $H_0=70$~km~s$^{-1}$~Mpc$^{-1}$.

\begin{figure*}[t]
\gridline{\fig{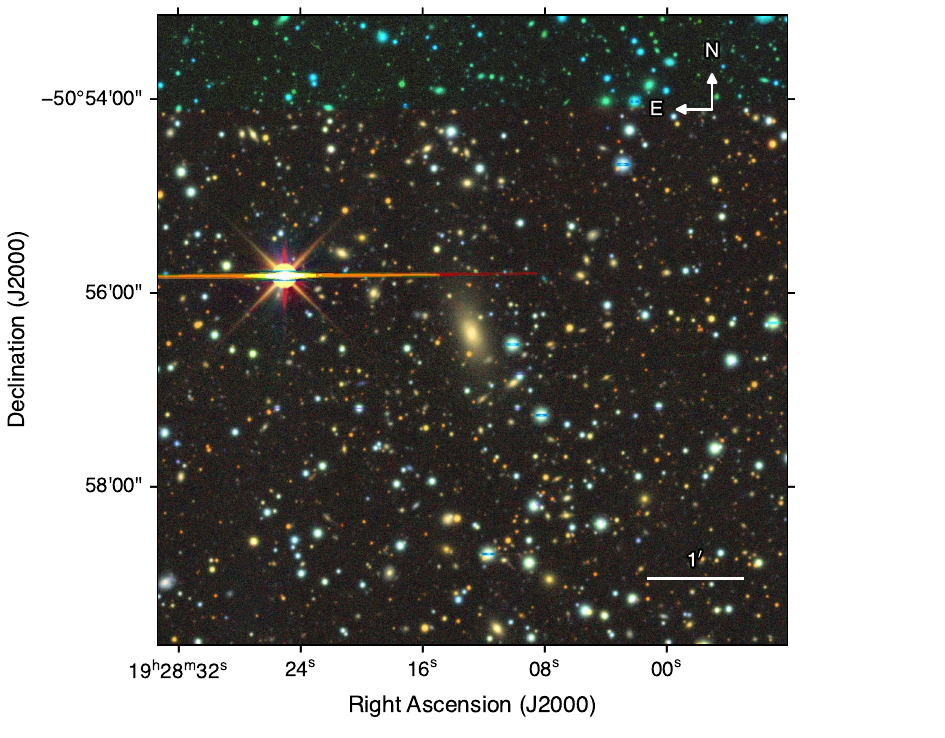}{0.47\textwidth}{(a)}
          \fig{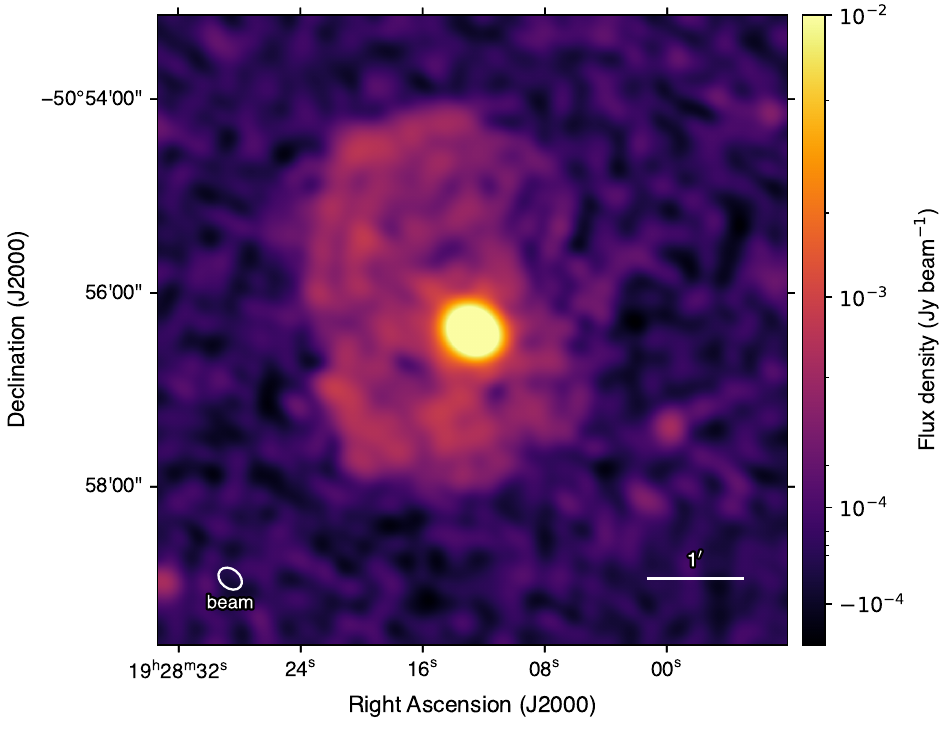}{0.47\textwidth}{(b)}}
\gridline{\fig{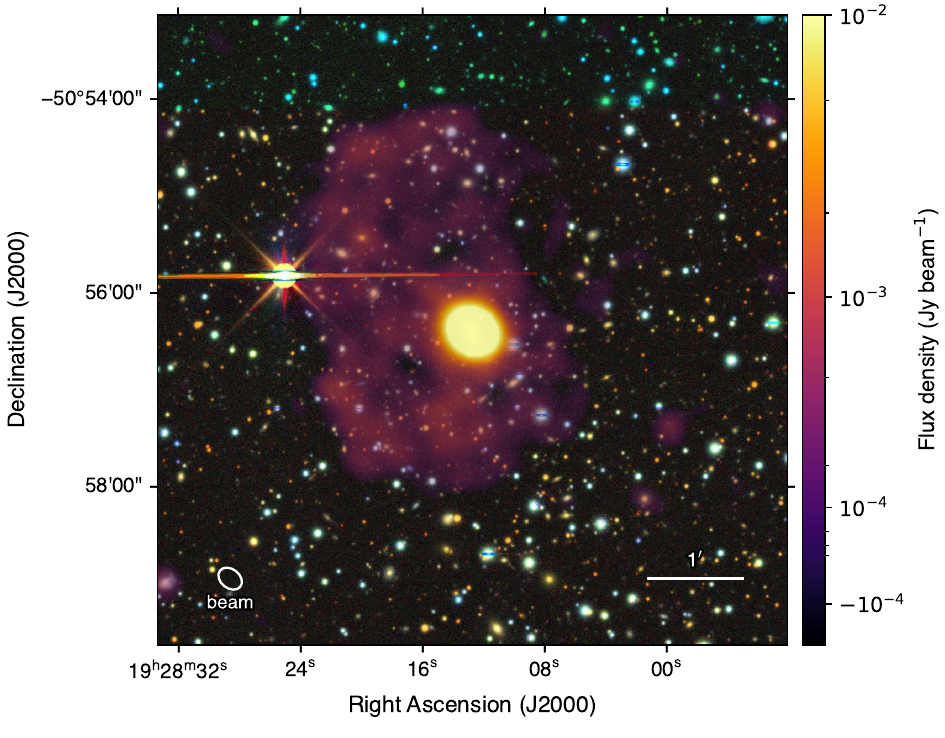}{0.47\textwidth}{(c)}
          \fig{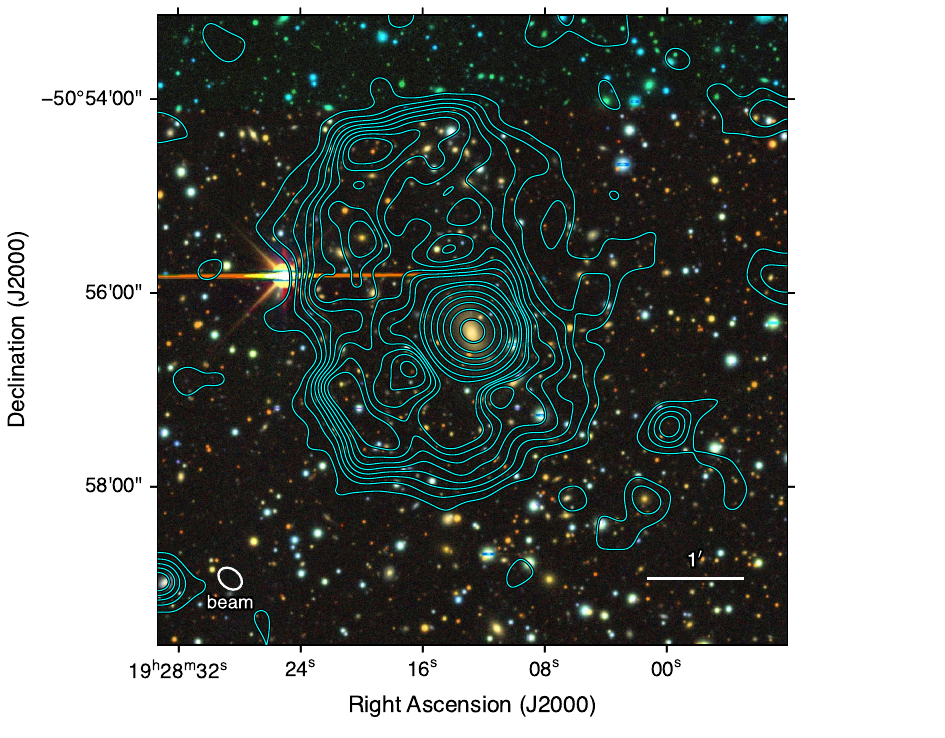}{0.47\textwidth}{(d)}}
\caption{ORC~J1928$-$5056 and its host LEDA~140850. (a) DESI $grz$ color image \citep{dey2019}; LEDA~140850 is the bright elliptical galaxy just below the center of the ORC. (b) ASKAP FLASH 856~MHz image of the same field. (c) Radio emission from (b) overlaid in color on the DESI image. (d) Radio contours at $(1,2,3,4,5,6,7,8,12,16,32,64,128,256,512)\times\sigma$, with $\sigma=70$~$\mu$Jy~beam$^{-1}$, drawn on the DESI image. The synthesized beam ($16.1^{\prime\prime}\times11.8^{\prime\prime}$) and a $1^{\prime}$ scale bar are shown in the corners of the panels. The ORC emission spans $\sim4^{\prime}$, or 630~kpc, at $z=0.1507$.}
\label{fig:four}
\end{figure*}

\section{Observations and Data Reduction}
\label{sec:obs}

ASKAP \citep{hotan2021} is a 36-antenna interferometer equipped with phased-array feeds that images $\sim$30~deg$^2$ instantaneously. The First Large Absorption Survey in H\,{\sc i} (FLASH; \citealp{allison2022}) is a wide-field survey for 21-cm absorption at 711.5--999.5~MHz with 18.5~kHz spectral channels and $\sim$2~hr integrations per field; its data products also include continuum images with a typical noise of $\sim$93~$\mu$Jy~beam$^{-1}$, about three times deeper than the Rapid ASKAP Continuum Survey (RACS; \citealp{mcconnell2020}).

The field containing the source was observed on 2026 January 28 as part of scheduling block 81516. The image is the standard continuum product. The synthesized beam is $16.1^{\prime\prime}\times11.8^{\prime\prime}$ at a position angle of $51.3\arcdeg$. The rms noise around the source is 70~$\mu$Jy~beam$^{-1}$.

We supplemented the FLASH data with the RACS low-, mid-, and high-band images (888, 1368, and 1656~MHz) \citep{mcconnell2020,hale2021,duchesne2024,duchesne2025} in order to measure the spectrum of the compact emission from the host galaxy.

\section{Results and Discussion}
\label{sec:results}

\subsection{Detection of a giant ORC}
\label{sec:det}

Figure~\ref{fig:four} presents the discovery image of the ORC. The radio image reveals a faint, diffuse, mildly elliptical structure, ORC~J1928$-$5056, centered on $(\alpha,\delta)_{\rm J2000}\approx(19^{\rm h}28^{\rm m}15^{\rm s},-50\degr56^{\prime}02^{\prime\prime})$ and enclosing the bright radio source associated with LEDA~140850. The major axis is $\sim4^{\prime}$ ($630$~kpc) and the minor axis is $\sim2.8^{\prime}$ ($440$~kpc) at $z=0.1507$.

To measure the flux density of the ORC, we integrated the emission within the polygon that encloses the source ($12.2$~arcmin$^2$) and subtracted the host galaxy, measured within its own polygon ($0.27$~arcmin$^2$). The total flux density is $193\pm10$~mJy, of which the host contributes $120\pm6.0$~mJy at 856~MHz (Section~\ref{sec:host}), leaving a flux density of $73\pm4$~mJy for the diffuse emission; the uncertainties are taken as 5\% calibration error. Dividing the diffuse flux density by the corresponding area gives a mean surface brightness of $6.2$~mJy~arcmin$^{-2}$, i.e.\ $1.7$~$\mu$Jy~arcsec$^{-2}$.

The observed luminosity of the diffuse emission at 856~MHz is $\approx4.5\times10^{24}$~W~Hz$^{-1}$. Adopting a spectral index of $\alpha=-2$ ($S_\nu\propto\nu^{\alpha}$, see Section~\ref{sec:host}), this corresponds to a rest-frame 1.4~GHz luminosity of $L_{1.4}=2.0\times10^{24}$~W~Hz$^{-1}$. ORC~J1928$-$5056 is therefore among the largest single ORCs known (Section~\ref{sec:compare}) and, once its host ($\alpha=-0.88$) is included, the most luminous ORC system known, with $L_{1.4}=6.6\times10^{24}$~W~Hz$^{-1}$.

\subsection{Host Galaxy Properties}
\label{sec:host}

The DESI color image (Figure~\ref{fig:four}a) shows that the ORC is centered close to LEDA~140850, a bright elliptical galaxy. LEDA~140850 is the BCG of Abell~3639, a richness class 2 cluster of Bautz--Morgan type I \citep{abell1989}. The ten cluster members have a mean redshift of $z=0.1495$ and a velocity dispersion of $\sigma_v=630\pm150$~km~s$^{-1}$ \citep{coziol2009}. The galaxy itself has a spectroscopic redshift $z=0.1507$, corresponding to a rest-frame velocity offset of $\approx310$~km~s$^{-1}$ ($\approx0.5\sigma_v$) from the cluster mean \citep{garilli1991}.

We compiled photometry of LEDA~140850 from the GALEX \citep{bianchi2017}, DESI \citep{dey2019}, 2MASS \citep{skrutskie2006}, and WISE \citep{wright2010} surveys; the measurements are listed in Table~\ref{tab:phot}. The colors, $g-r=1.062$, $i-z=0.255$,
$J-K_s=0.931$, indicate an early-type galaxy \citep{janz2009,jarrett2003},
and $W1-W2=0.209$ lies well below the threshold of 0.8 usually adopted to select AGNs \citep{stern2012}, implying that no AGN is currently active. The $W2-W3$ color (2.077~mag) is redder than the typical of passive early-type galaxies, but this is expected given circumstellar dust around asymptotic giant branch stars, which dominates the 12~$\mu$m emission of quiescent early-type galaxies \citep{bressan1998,athey2002,villaume2015}.

\begin{figure*}
    \centering
 \includegraphics[width=0.9\textwidth]{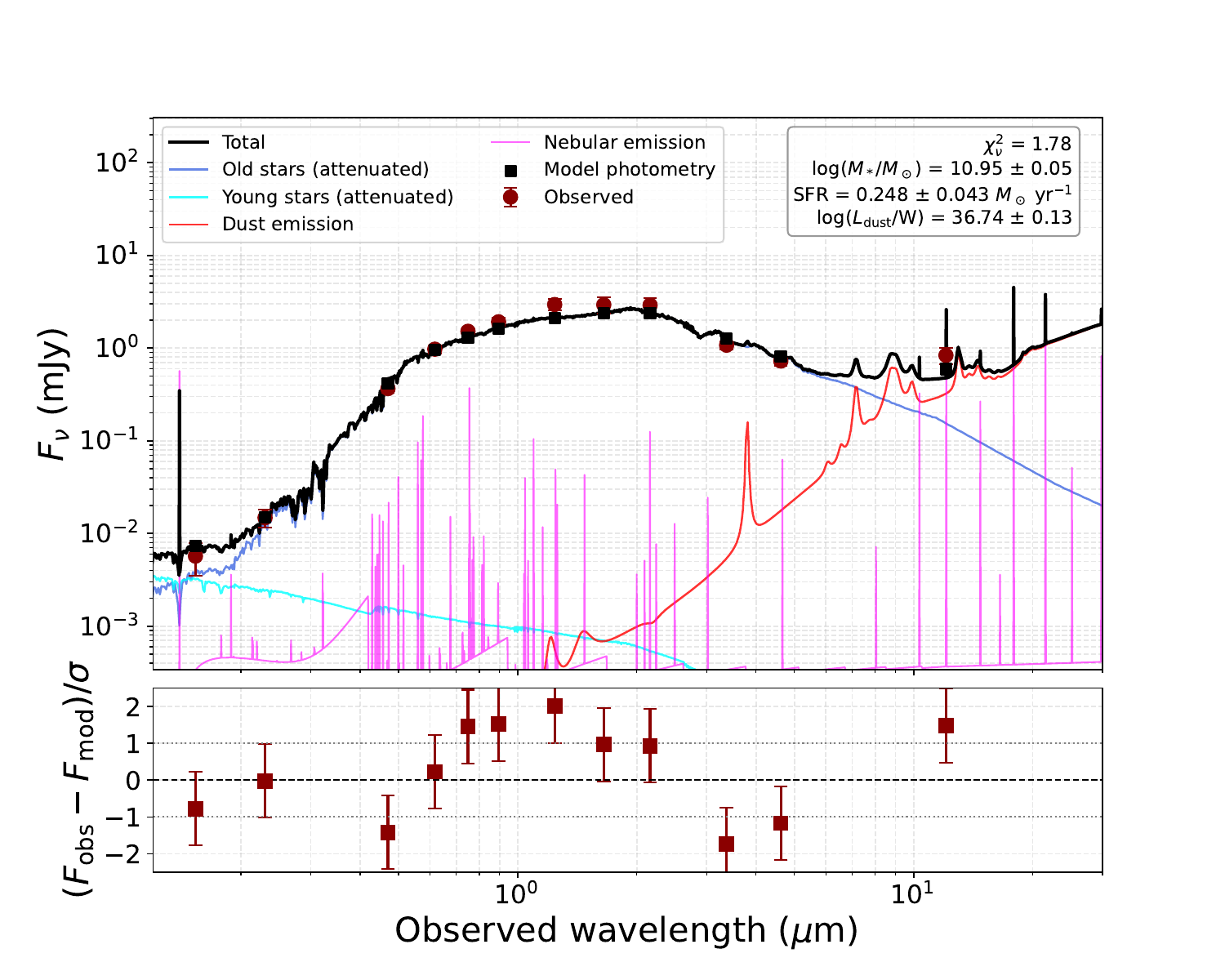}
    \caption{\textsc{cigale} SED fit of LEDA~140850. The best fit gives
    $\log(M_{\ast}/M_{\odot})=10.95\pm0.05$ and an SFR of $0.248\pm0.043$~$M_{\odot}$~yr$^{-1}$.}
    \label{fig:sed}
\end{figure*}

\begin{figure}[t]
    \centering
    \includegraphics[width=0.9\columnwidth]{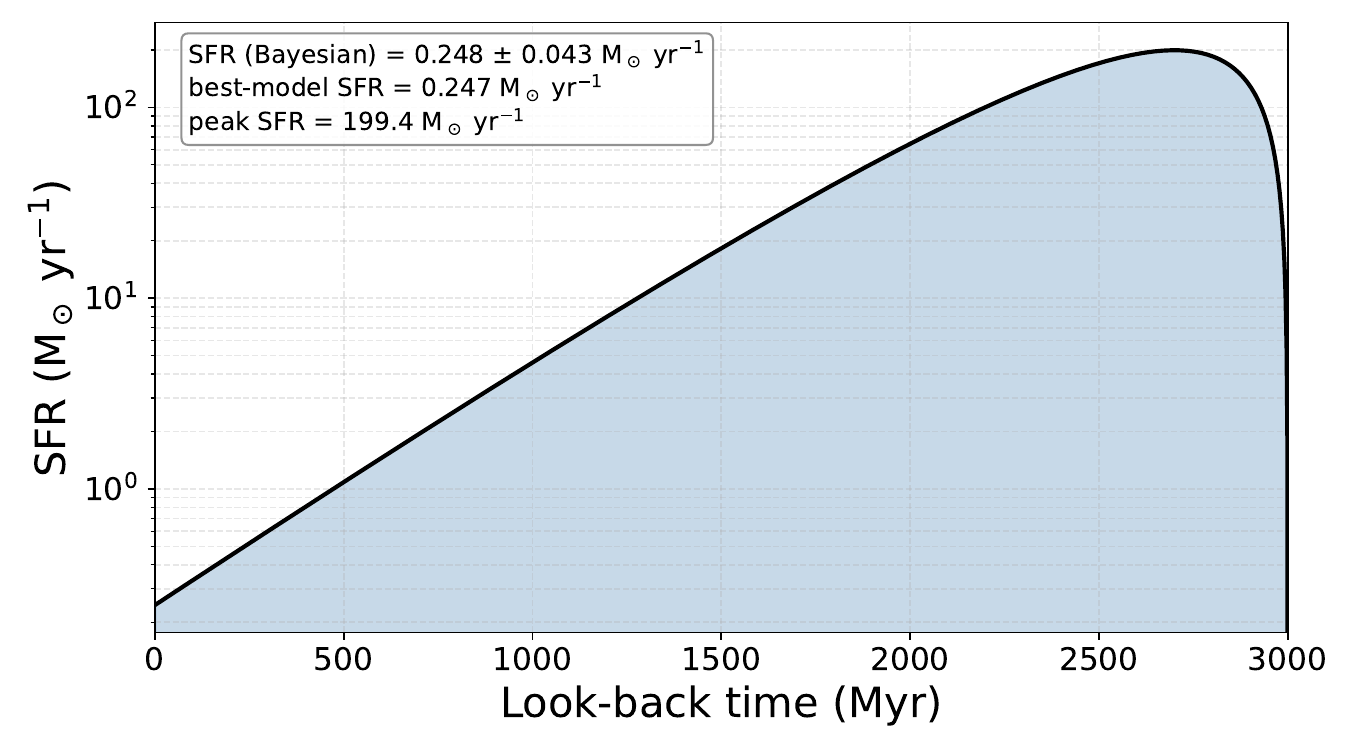}
    \caption{The star formation history of LEDA~140850.}
    \label{fig:sfh}
\end{figure}

To derive the stellar mass and star formation rate (SFR), we fitted the spectral energy distribution (SED; Table~\ref{tab:phot}) with
\textsc{cigale} \citep{boquien2019}, assuming a delayed
exponential star-formation history. The results are shown in Figures~\ref{fig:sed} and~\ref{fig:sfh}. We obtain a stellar mass of $\log(M_{\ast}/M_{\odot})=10.95\pm0.05$ and an SFR of $0.248\pm0.043$~$M_{\odot}$~yr$^{-1}$, confirming that the host galaxy is a massive, quenched galaxy.

Given such a low SFR, its contribution to the radio emission of the host is negligible \citep{kennicutt2012}. The radio emission must therefore be powered by an AGN, and a radio source of this luminosity in a passive elliptical galaxy is expected to take the form of jets. The optical spectrum shows no obvious evidence of ongoing AGN activity \citep{garilli1991}, consistent with the WISE color $W1-W2=0.209$. Consequently, the AGN is no longer active and the radio emission comes from the relic jet. 

\begin{figure}[t]
\includegraphics[width=0.9\columnwidth]{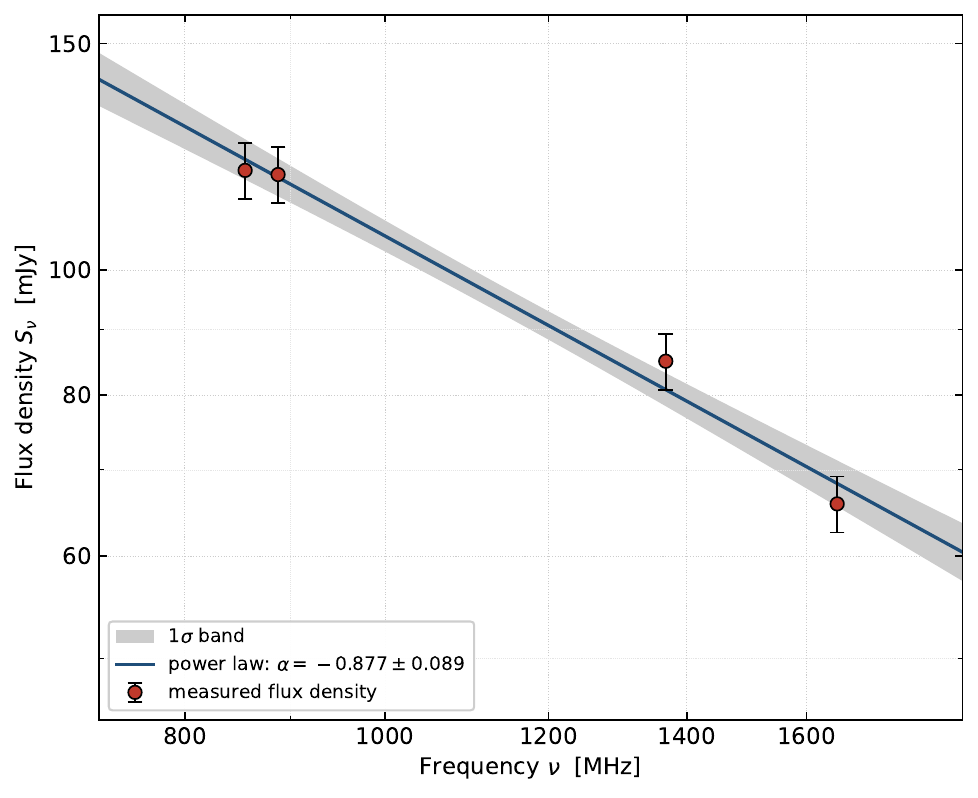}
\caption{Radio spectrum of the host galaxy LEDA~140850, combining the FLASH 856~MHz measurement with the RACS low-, mid-, and high-band measurements (888, 1368, 1656~MHz). The diffuse ORC emission is only marginally detected in the RACS low band, and all flux densities shown refer to the host galaxy. A single power law with $\alpha=-0.877\pm0.089$ fits all four data points, and the shaded band indicates the $1\sigma$ uncertainty. Such a steep index indicates optically thin synchrotron emission from an aged electron population, as expected for AGN jets.}
\label{fig:sed_radio}
\end{figure}

The radio SED constructed from the ASKAP measurements (Figure~\ref{fig:sed_radio}) is a clean power law with $\alpha=-0.877\pm0.089$ between 856 and 1656~MHz. Such a steep, optically thin synchrotron spectrum is characteristic of jet and lobe emission. We also retrieved low-frequency radio data from TGSS \citep{intema2017} and GLEAM \citep{hurleywalker2017}. TGSS detects a flux density of 2.7~Jy at 150~MHz, whereas the host spectrum alone extrapolates to only 0.55~Jy at that frequency, leaving an excess of 2.2~Jy. GLEAM, at 200~MHz, gives 1.9~Jy, against 0.43~Jy extrapolated for the host. If both excesses are ascribed to the diffuse ORC, they imply $\alpha=-1.9$ and $-2.1$ from TGSS and GLEAM, respectively, i.e.\ an extremely steep spectrum.

\begin{deluxetable}{lccc}
\tablecaption{Multi-band Photometry of LEDA~140850.\label{tab:phot}}
\tablehead{
\colhead{Band} & \colhead{Catalog} & \colhead{System} &
\colhead{$m$ (mag)} 
}
\startdata
FUV & GALEX & AB & $22.017\pm0.407$ \\
NUV & GALEX & AB & $20.974\pm0.210$ \\
g & DESI & AB & $17.492\pm0.001$ \\
r & DESI & AB & $16.430\pm0.002$ \\
i & DESI & AB & $15.947\pm0.002$ \\
z & DESI & AB & $15.692\pm0.003$ \\
J & 2MASS & Vega & $14.331\pm0.108$ \\
H & 2MASS & Vega & $13.857\pm0.183$ \\
Ks & 2MASS & Vega & $13.400\pm0.180$ \\
W1 & WISE & Vega & $13.639\pm0.029$ \\
W2 & WISE & Vega & $13.430\pm0.038$ \\
W3 & WISE & Vega & $11.353\pm0.182$ \\
\enddata
\tablecomments{$m$ is corrected for Galactic
extinction with $E(B-V)=0.053$ \citep{schlafly2011}.}
\end{deluxetable}

\begin{deluxetable*}{lcccccc}
\tablecaption{Comparison of ORC J1928$-$5056 with known ORCs.\label{tab:orcs}}
\tablehead{
\colhead{Name} & \colhead{$z$} & \colhead{Diameter} & \colhead{$S_{\rm int}$} &
\colhead{$\nu$} & \colhead{Ref.} & \colhead{$L_{1.4}^{\rm rest}$} \\
\colhead{} & \colhead{} & \colhead{(kpc)} & \colhead{(mJy)} &
\colhead{(GHz)} & \colhead{} & \colhead{($10^{24}$ W Hz$^{-1}$)}
}
\startdata
ORC1 J2103$-$6200\tablenotemark{a} & 0.551 & 520 & $6.26\pm1.25$ & 0.944 &1, 2 & $5.2\pm1.0$ \\
ORC4 J1555+2726\tablenotemark{b} & 0.451 & 440 & $5.3\pm0.7$ & 1.4 & 1, 6 & $3.9\pm0.6$ \\
ORC5 J0102$-$2450\tablenotemark{c} & 0.27 & 300 & $3.9\pm0.2$ & 0.944 & 3 & $0.61\pm0.04$ \\
ORC6 J1841$-$6547\tablenotemark{d} & 0.18 & $2\times720$ & $14.7\pm0.5$ & 0.944 & 4 & $0.60\pm0.02$ \\
Cloverleaf ORC\tablenotemark{e} & 0.046 & 227 & \nodata & \nodata & 5 & \nodata \\
MIGHTEE ORC J0219$-$0505\tablenotemark{f} & 0.196 & 114 & $1.2\pm0.1$ & 1.2 & 7 & $0.11\pm0.01$ \\
Bubble ORC J0949$-$8553 \tablenotemark{g} & 0.040 & $2\times228$ & $56.8\pm2.9$ & 0.944 & 8 & $0.14\pm0.01$ \\
ORC J1928$-$5056\tablenotemark{h} & 0.1507 & 630 & $193\pm 10$ & 0.86 & 9 & $6.6\pm 0.3$ \\
\enddata
\tablecomments{To calculate the rest-frame 1.4~GHz luminosity, the adopted spectral indices: ORC1 $\alpha=-1.17$;
ORC4 $\alpha=-0.92$; ORC5 $\alpha=-0.8$; ORC6 host $\alpha=-1$, two shells $\alpha=-2.8$ and $\alpha=-3.4$; MIGHTEE $\alpha=-1$ (assumed; its host has a measured $-1.22\pm0.11$); ORC J1928$-$5056 host $\alpha=-0.88$, diffuse shell $\alpha=-2$.}

\tablenotetext{a}{The central galaxy ORC 1 is detected as a compact radio source with MeerKAT ($0.091$~mJy at 1284~MHz).}
\tablenotetext{b}{The total flux density includes the faint compact central source.}
\tablenotetext{c}{A faint compact radio source associated with the central elliptical host is present; the quoted flux density is dominated by the ring. The flux density uncertainty is taken as 5\% of the total flux density.}
\tablenotetext{d}{The total flux density includes the compact central source; the shells alone have $\sim13.4$~mJy. }
\tablenotetext{e}{A central AGN is present (X-ray and radio peak at the central galaxy), but no integrated radio flux density has been published.}
\tablenotetext{f}{The flux density includes the compact central source associated with the host ($0.171\pm0.018$~mJy, likely an AGN).}
\tablenotetext{g}{The total flux density includes the faint compact central source.}
\tablenotetext{h}{Total flux density and luminosity. The bright central radio source of the host galaxy ($\approx120$~mJy) is included.}
\tablerefs{(1) \citep{norris2021}; (2) \citep{norris2022}; (3) \citep{koribalski2021}; (4) \citep{koribalski2026}; (5) \citep{bulbul2024}; (6) \citep{rupke2024}; (7) \citep{norris2025}; (8) \citep{su2026}; (9) this work.}
\end{deluxetable*}

\subsection{Comparison with Known ORCs}
\label{sec:compare}

Table~\ref{tab:orcs} places ORC~J1928$-$5056 in the context of the known population. Confirmed single ORCs span 100--500~kpc in diameter with flux densities of several mJy; see also the five sources cataloged by \citet{gupta2025}. ORC6 has a higher flux density, although it consists of two intersecting shells of 720~kpc diameter \citep{koribalski2026}. Recently, the ORC system ORC~J0949$-$8553, which contains two bubbles of 228~kpc diameter, was reported to have a flux density as high as 60~mJy \citep{su2026}. With an extent of 630~kpc, ORC~J1928$-$5056 is the largest single ORC known, and with 193~mJy at 856~MHz (including 120~mJy from the host galaxy) it is the brightest ORC system known. Its diffuse emission luminosity at 1.4 GHz in the rest frame, $L_{1.4}=2\times10^{24}$~W~Hz$^{-1}$, is comparable to those of the two most luminous previously known ORCs ($5.2\times10^{24}$~W~Hz$^{-1}$ for ORC1 and $3.9\times10^{24}$~W~Hz$^{-1}$ for ORC4; Table~\ref{tab:orcs}). For the system as a whole, $L_{1.4}=6.6\times10^{24}$~W~Hz$^{-1}$, which is higher than that of any ORC reported so far. Comparing with the known ORCs in Table~\ref{tab:orcs}, we find that all of them show compact radio emission associated with their host galaxies, suggesting that nuclear radio emission may be essential to the formation of ORCs.

The host of our ORC, LEDA~140850, is a massive, quiescent elliptical galaxy, closely resembling the hosts of confirmed ORCs, although it is far more radio-loud than they are. To our knowledge, LEDA~140850 is also the first ORC host that is the BCG of a richness class 2 cluster; previously, ORC-like shells had been associated only with galaxy groups or dense environments \citep{norris2021b,bulbul2024,koribalski2024b}. The combination of a cluster environment (Section~\ref{sec:models}) and a currently radio-loud AGN at its center makes this system a valuable test case for models linking ORCs to AGN activity and intracluster shocks.

\subsection{On the Models of ORCs}
\label{sec:models}

We now compare the main proposed mechanisms with the observed morphology, size, flux density, spectrum, and host galaxy of ORC~J1928$-$5056.

Cataclysmic explosion models, in which the diffuse shell is a blast wave from  tidal disruption event \citep{omar2022b}, are strongly disfavored. The shell has a radius of 315~kpc, so the medium into which it expands is the intracluster medium (ICM) of Abell~3639.

We estimate the ICM pressure from scaling relations. Using the relation between galaxy velocity dispersion and gas temperature, $kT=\mu m_p\sigma_v^2$ \citep[e.g.,][]{lovisari2021}, with $\sigma_v=630$~km~s$^{-1}$ and mean molecular weight $\mu=0.6$, gives $kT\approx2.5$~keV. At a radius of $\sim300$~kpc, the typical gas density is $n_e\sim10^{-3}$~cm$^{-3}$ in the REXCESS cluster sample \citep{croston2008}. With this value, we obtain $P\sim2\,n_ekT=8\times10^{-12}$~dyn~cm$^{-2}$. Inflating a shell of this radius requires an energy $E\sim(5/2)PV\approx8\times10^{61}$~erg, corresponding to a relativistic enthalpy of $4PV\sim1.2\times10^{62}$~erg \citep{birzan2004,forman2005}, where $V$ is the volume enclosed by the shell of radius 315~kpc. This energy far exceeds what  tidal disruption event can supply. Starburst-driven wind bubbles \citep{lochhaas2018,fujita2024} are disfavored as well given the negligible SFR of the host inferred from the SED fitting in Section \ref{sec:host}.

Virial-shock models \citep{yamasaki2024} interpret ORCs as accretion shocks around galaxy halos of $\sim10^{13}\,M_\odot$. The shock radius is $r_{\rm sh}=f_{\rm sh}r_{\rm vir}$, where $f_{\rm sh}$ is a shock geometry parameter in the range $0.6$--$1.2$, and $r_{\rm vir}$ is the virial radius ($\sim400$~kpc) for a $\sim10^{13}\,M_\odot$ galaxy halo. The shock radius is therefore expected to be $\sim240$--$480$~kpc, comparable to the size of our ORC. Moreover, the virial-shock model predicts observed radio flux densities of $\sim$mJy at 1~GHz for sources at $z\sim0.5$. This corresponds to a luminosity consistent with our observed $L_{1.4}=2\times10^{24}$~W~Hz$^{-1}$. In terms of both size and energy, the virial-shock model is a plausible mechanism for the ORC we observe.

A merger-shock origin \citep{dolag2023,ivleva2026} also deserves consideration. The host, LEDA~140850, is a massive, elliptical, quenched galaxy, as expected in simulations once the merger shock reaches the size of observed ORCs, i.e.\ several hundred kpc \citep{dolag2023}. The merger-shock scenario has been invoked for several known ORCs \citep[e.g.,][]{koribalski2024,koribalski2026,bulbul2024}. Although the required energy, $\sim10^{62}$~erg, could be supplied by a galaxy merger \citep{dolag2023,ivleva2026}, the radio flux densities observed for previously known ORCs are difficult to reconcile with the theoretical predictions. The radio luminosity of the diffuse shell of our ORC, $L_{1.4}=2\times10^{24}$~W~Hz$^{-1}$, lies at the upper end of the range spanned by previously known ORCs (Table~\ref{tab:orcs}). The discrepancy between observed and predicted radio luminosities therefore applies to our ORC as well.

AGN-related scenarios are more natural. The host is radio-loud, and the ultra-steep spectral index ($\alpha\approx-2$) is the signature of aged or re-accelerated synchrotron plasma. The diffuse shell could be a jet-inflated bubble or a lobe pair seen in projection \citep{lin2024}, or a ``phoenix'' --- fossil lobes re-energized by a passing merger shock \citep{shabala2024}. The shell energy is $\sim10^{62}$~erg, which is within the output of a powerful radio galaxy (jet power $\approx10^{45}$--$10^{46}$~erg~s$^{-1}$) over a few $10^{8}$~yr, comparable to the sound-crossing time of the shell ($\approx3\times10^{8}$~yr). Moreover, lobe models can naturally account for the offset between the host galaxy and the center of the shell.

\subsection{Future Observations}
\label{sec:future}

Several kinds of follow-up observations will provide further information on the ORC. (i) Deep broadband imaging, including polarimetry, with MeerKAT will reveal the internal structure and the strength and geometry of the magnetic field, which are crucial for understanding the nature of ORCs \citep[e.g.,][]{norris2022,taziaux2025}. (ii) A spatially resolved spectral index map will be important: a uniform ultra-steep spectrum would support a single aged plasma volume, whereas localized flattening would indicate re-acceleration by shocks \citep[e.g.,][]{su2026}. (iii) Optical integral-field spectroscopy will map the ionized gas in the host galaxy and within the shell, testing whether line emission such as [O\,{\sc ii}] from LEDA~140850 extends on large scales, as it does in the host of ORC~4 \citep{coil2024}, and will help to distinguish between the driving mechanisms.

\section{Summary}
\label{sec:summary}

We have reported the discovery of ORC~J1928$-$5056, a giant odd radio circle found in 856~MHz ASKAP FLASH continuum data, and have analyzed its properties and those of its host galaxy. Our main conclusions are as follows.

\begin{enumerate}
\setlength{\itemsep}{2pt plus 0pt}
\setlength{\parsep}{0pt plus 0pt}
\setlength{\topsep}{4pt plus 0pt}
\item ORC~J1928$-$5056 has a faint and nearly circular shell of $\sim4^{\prime}$ (630~kpc) in diameter, centered near LEDA~140850 at $z=0.1507$; it is the largest single ORC known, extending the size range of the population.

\item The diffuse shell has a flux density of $73\pm4$~mJy, excluding the host emission, and a mean surface brightness of $6.2$~mJy~arcmin$^{-2}$, giving $L_{1.4}=2\times10^{24}$~W~Hz$^{-1}$, comparable to those of the most luminous
known ORCs. Including the host emission, the system has $L_{1.4}=6.6\times10^{24}$~W~Hz$^{-1}$, making it the most luminous ORC reported to date.

\item The host is the BCG of the richness class 2 cluster Abell~3639: a massive ($M_\ast\sim10^{11}\,M_\odot$) and passive elliptical galaxy whose radio luminosity far exceeds that expected from star formation, implying the presence of an AGN radio jet with a steep spectrum ($\alpha=-0.877\pm0.089$).

\item AGN lobes, possibly re-energized by a shock, provide the most natural explanation of the ORC. Stellar explosions and starburst winds are disfavored. Deep broadband radio imaging, spectral-index and polarization mapping, and optical integral-field spectroscopy will help to reveal the nature of the ORC.
\end{enumerate}

\section*{Acknowledgments}

This work is supported by the National Science and Technology Major Project of China (No. 2024ZD1100601), the National SKA Program of China (No. 2025SKA0130100). RZS acknowledges the support from the National Science Foundation of China (grant 12603034), the Shanghai Super Postdoctoral Incentive Program (No. 2025322), the China Postdoctoral Science Foundation (Grant No. 2025M783232). MFG is supported by the National Science Foundation of China (grant 12473019), the National SKA Program of China (Grant No. 2022SKA0120102), the Shanghai Pilot Program for Basic Research-Chinese Academy of Science, Shanghai Branch (JCYJ-SHFY-2021-013), and the China Manned Space Project with No. CMS-CSST-2025-A07.

The Australian SKA Pathfinder is part of the Australia Telescope
National Facility (https://ror.org/05qajvd42), which is managed by CSIRO. Operation of ASKAP is funded by the Australian Government with support from the National Collaborative Research Infrastructure Strategy. ASKAP uses the resources of the Pawsey Supercomputing Centre. Establishment of ASKAP, the Murchison Radio-astronomy Observatory, and the Pawsey Supercomputing Centre are initiatives of the Australian Government, with support from the Government of Western Australia and the Science and Industry Endowment Fund. We acknowledge the Wajarri Yamatji people as the traditional owners of the Observatory site.

The Legacy Surveys consist of three individual and complementary projects: the Dark Energy Camera Legacy Survey (DECaLS; Proposal ID \#2014B-0404; PIs: David Schlegel and Arjun Dey), the Beijing-Arizona Sky Survey (BASS; NOAO Prop. ID \#2015A-0801; PIs: Zhou Xu and Xiaohui Fan), and the Mayall z-band Legacy Survey (MzLS; Prop. ID \#2016A-0453; PI: Arjun Dey). DECaLS, BASS and MzLS together include data obtained, respectively, at the Blanco telescope, Cerro Tololo Inter-American Observatory, NSF's NOIRLab; the Bok telescope, Steward Observatory, University of Arizona; and the Mayall telescope, Kitt Peak National Observatory, NOIRLab. Pipeline processing and analyses of the data were supported by NOIRLab and the Lawrence Berkeley National Laboratory (LBNL). The Legacy Surveys project is honored to be permitted to conduct astronomical research on Iolkam Du'ag (Kitt Peak), a mountain with particular significance to the Tohono O'odham Nation.

NOIRLab is operated by the Association of Universities for Research in Astronomy (AURA) under a cooperative agreement with the National Science Foundation. LBNL is managed by the Regents of the University of California under contract to the U.S. Department of Energy.

This project used data obtained with the Dark Energy Camera (DECam), which was constructed by the Dark Energy Survey (DES) collaboration. Funding for the DES Projects has been provided by the U.S. Department of Energy, the U.S. National Science Foundation, the Ministry of Science and Education of Spain, the Science and Technology Facilities Council of the United Kingdom, the Higher Education Funding Council for England, the National Center for Supercomputing Applications at the University of Illinois at Urbana-Champaign, the Kavli Institute of Cosmological Physics at the University of Chicago, Center for Cosmology and Astro-Particle Physics at the Ohio State University, the Mitchell Institute for Fundamental Physics and Astronomy at Texas A\&M University, Financiadora de Estudos e Projetos, Fundacao Carlos Chagas Filho de Amparo, Financiadora de Estudos e Projetos, Fundacao Carlos Chagas Filho de Amparo a Pesquisa do Estado do Rio de Janeiro, Conselho Nacional de Desenvolvimento Cientifico e Tecnologico and the Ministerio da Ciencia, Tecnologia e Inovacao, the Deutsche Forschungsgemeinschaft and the Collaborating Institutions in the Dark Energy Survey. The Collaborating Institutions are Argonne National Laboratory, the University of California at Santa Cruz, the University of Cambridge, Centro de Investigaciones Energeticas, Medioambientales y Tecnologicas-Madrid, the University of Chicago, University College London, the DES-Brazil Consortium, the University of Edinburgh, the Eidgenossische Technische Hochschule (ETH) Zurich, Fermi National Accelerator Laboratory, the University of Illinois at Urbana-Champaign, the Institut de Ciencies de l’Espai (IEEC/CSIC), the Institut de Fisica d’Altes Energies, Lawrence Berkeley National Laboratory, the Ludwig Maximilians Universitat Munchen and the associated Excellence Cluster Universe, the University of Michigan, NSF’s NOIRLab, the University of Nottingham, the Ohio State University, the University of Pennsylvania, the University of Portsmouth, SLAC National Accelerator Laboratory, Stanford University, the University of Sussex, and Texas A\&M University.

BASS is a key project of the Telescope Access Program (TAP), which has been funded by the National Astronomical Observatories of China, the Chinese Academy of Sciences (the Strategic Priority Research Program “The Emergence of Cosmological Structures” Grant \# XDB09000000), and the Special Fund for Astronomy from the Ministry of Finance. The BASS is also supported by the External Cooperation Program of Chinese Academy of Sciences (Grant \# 114A11KYSB20160057), and Chinese National Natural Science Foundation (Grant \# 12120101003, \# 11433005).

The Legacy Survey team makes use of data products from the Near-Earth Object Wide-field Infrared Survey Explorer (NEOWISE), which is a project of the Jet Propulsion Laboratory/California Institute of Technology. NEOWISE is funded by the National Aeronautics and Space Administration.

The Legacy Surveys imaging of the DESI footprint is supported by the Director, Office of Science, Office of High Energy Physics of the U.S. Department of Energy under Contract No. DE-AC02-05CH1123, by the National Energy Research Scientific Computing Center, a DOE Office of Science User Facility under the same contract; and by the U.S. National Science Foundation, Division of Astronomical Sciences under Contract No. AST-0950945 to NOAO.


%





\bibliography{orc}{}
\bibliographystyle{aasjournal}


\end{CJK*}
\end{document}